\documentclass{appolb}
\usepackage{graphicx}
\usepackage{booktabs}
\usepackage[italic]{hepnames}
\usepackage{amsfonts}
\usepackage{amsthm}
\usepackage{bm}
\usepackage{slashed}
\usepackage{dsfont}
\usepackage{tikz}
\usepackage[dvipsnames]{xcolor}
\usepackage[shortlabels]{enumitem}
\usepackage{xspace}
\usepackage{siunitx}
\usepackage{bookmark}
\usepackage{bbm}
\usepackage{subcaption}

\def\1eq#1{Eq.\nobreak\thinspace(\ref{#1})}

\def\2eqs#1#2{Eqs.\nobreak\thinspace(\ref{#1}) and\nobreak\thinspace(\ref{#2})}
\def\3eqs#1#2#3{Eqs.\nobreak\thinspace(\ref{#1}),\nobreak\thinspace(\ref{#2}) and\nobreak\thinspace(\ref{#3})}

\def\fig#1{\hyperref[#1]{Fig.\nobreak\thinspace\ref*{#1}}}
\def\figA#1{\hyperref[#1]{Fig.\nobreak\thinspace\ref*{#1}A}}
\def\figB#1{\hyperref[#1]{Fig.\nobreak\thinspace\ref*{#1}B}}
\def\figC#1{\hyperref[#1]{Fig.\nobreak\thinspace\ref*{#1}C}}
\def\tab#1{\hyperref[#1]{Tab.\nobreak\thinspace\ref*{#1}}}

\def\sect#1{\hyperref[#1]{Sec.\nobreak\thinspace\ref*{#1}}}
\def\appref#1{\hyperref[#1]{App.\nobreak\thinspace\ref*{#1}}}

\def\eg{{\it e.g.}, }

\newcommand{\be}{\begin{equation}}
\newcommand{\ee}{\end{equation}}

\newcommand{\bea}{\begin{eqnarray}}
\newcommand{\eea}{\end{eqnarray}}

\def\is{S^{-1}}             

\def\g{\Gamma}              

\def\ga{\Gamma_{\!5}}

\begin{document} 
\title{Pions reloaded%
\thanks{Presented by J.P. at the Excited QCD Workshop 2026, Granada, Spain}%
}
\author{M.N.~Ferreira
\address{\mbox{Instituto de Física, Universidade Federal do Rio Grande do Sul}, Caixa Postal 15051, 91501-970, Porto Alegre, RS, Brazil}
\\[3mm]
{A.S.~Miramontes, J.M.~Morgado, J.~Papavassiliou
\address{\mbox{Department of Theoretical Physics and IFIC, University of Valencia and CSIC}, E-46100, Valencia, Spain}}
\\[3mm]
J.M.~Pawlowski
\address{\mbox{Institut f\"ur Theoretische Physik, Universit\"at Heidelberg}, Philosophenweg 16, Heidelberg, 69120, Germany}
\address{\mbox{ExtreMe Matter Institute EMMI, GSI, Planckstrasse 1, Darmstadt, 64291, Germany}}
}

\maketitle
\begin{abstract}

We present a novel version of the pion Bethe-Salpeter equation in the chiral limit,
solved using as ingredients  
state-of-the-art QCD correlation 
functions. The constraints imposed by the axial 
Ward-Takahashi identities are exactly fulfilled,
both formally and numerically. 

\end{abstract}
  
\section{Introduction}

The 
wealth of information  
on the structure of the QCD correlation functions, 
amassed by functional 
methods and lattice simulations \cite{Ferreira:2023fva,Pawlowski:2005xe,Huber:2018ned,Kizilersu:2021jen}, 
may be profitably utilized    
in the physics of hadrons
only within sophisticated frameworks
that respect the constraints imposed  
by chiral symmetry. 
Recently, a theoretical approach was developed
in \cite{Miramontes:2025imd}, 
which allows for the 
self-consistent inclusion of 
dressed correlation functions in the dynamical equations describing the physics of mesons. 
This approach was judiciously simplified in
\cite{Ferreira:2025wpu}, leading to a rather tractable set of dynamical equations. 
In particular, the quark-gap  
and the pion Bethe Salpeter equation (BSE) 
admit fully-dressed quark-gluon vertices, denoted by $\Gamma_{\mu}(q,r,-p)$, 
containing all possible tensorial structures.  
In this presentation we showcase the application of this
method in the study of massless pions.

\section{Main ingredients} 

\begin{figure}[t]
    \hspace*{-0.75cm}
    \includegraphics[scale=1]{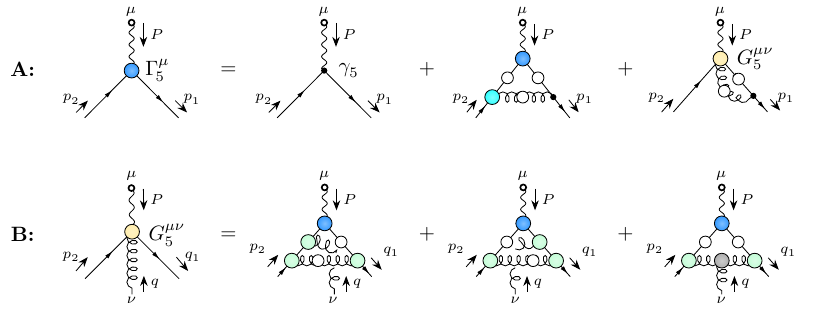}
    \caption{\textbf{\textit{Panel A:}} The exact SDE for the axial-vector $\ga^\mu$ . \textbf{\textit{Panel B:}} The SDE for the vertex $G_5^{\mu\nu}$, after implementing the SV approximation.}
    \label{fig:sdeg5}
\end{figure}

Our starting point is the axial-vector 
vertex $\ga^\mu(P,p_2,-p_1)$, which satisfies 
the Ward-Takahashi identity (WTI) \cite{Miransky:1994vk,Itzykson:1980rh} 
\be
\label{WTI1}
-P_\mu\ga^\mu(P,p_2,-p_1)=\is(p_1)\gamma_5+\gamma_5\is(p_2) \,,
\ee
where $S^{-1}(p)=A(p^2)\slashed{p}-B(p^2)$
is the inverse quark propagator.
The dynamical breaking of the chiral symmetry gives rise to a non-vanishing 
$B(p^2)$, and from \1eq{WTI1} one obtains:   
$\lim_{P\to 0}P_\mu \ga^\mu(P,p_2,-p_2)=2B(p^2)\gamma_5$.
This, in turn, forces 
$\ga^\mu(P,p_2,-p_1)$ to contain a 
longitudinally coupled 
massless pole,  $(P^\mu/P^2)\,\chi(P,p_2,-p_1)\gamma_5$, 
associated with the 
corresponding  
massless Goldstone  boson (pion). 
In the limit $P\to 0$, we have that 
\be
\chi(0,p,-p)=\chi_1(p^2)+\chi_3(p^2)\slashed{p}=2B(p^2) \,,
\ee
from which we get two key constraints
for the components of the pion BS amplitude, namely
\cite{Miransky:1994vk} 
\begin{align}
    \chi_1(p^2)&=2B(p^2)\,,& \chi_3(p^2)&=0\,.
\label{WTIcon}
\end{align}
The SDE governing $\ga^\mu(P,p_2,-p_1)$ 
is shown in \figA{fig:sdeg5} \cite{Bender:2002as}. The 
element crucial for effectuating  
the transition beyond the RL is the 
axial-vector-gluon vertex, $G_5^{\mu\nu}(P,q,p_2,-q_1)$, denoted by a yellow circle; 
it is composed of 
``one-loop-dressed" diagrams, which, at the 
level of $\ga^\mu(P,p_2,-p_1)$, correspond 
to ``two-loop dressed" contributions. 
$G_5^{\mu\nu}(P,q,p_2,-q_1)$ 
obeys the WTI \cite{Chang:2009zb}
\begin{align}
-iP_\mu G_5^{\mu\nu}(P,q,p_2,-q_1)=\g^\nu(q,p_1,-q_1)\gamma_5+\gamma_5\g^\nu(q,p_2,-q_2)\,,
\label{eq:G5WTI}
\end{align}
where $q_i:=p_i+q$, 
which guarantees that the diagrammatic representation of 
$\ga^\mu(P,p_2,-p_1)$, given by \figA{fig:sdeg5},
satisfies the WTI of \1eq{WTI1} exactly. 

\section{The symmetric-vertex approximation}

\begin{figure}[t]
    \hspace*{-2.5cm}
    \includegraphics[scale=1.25]{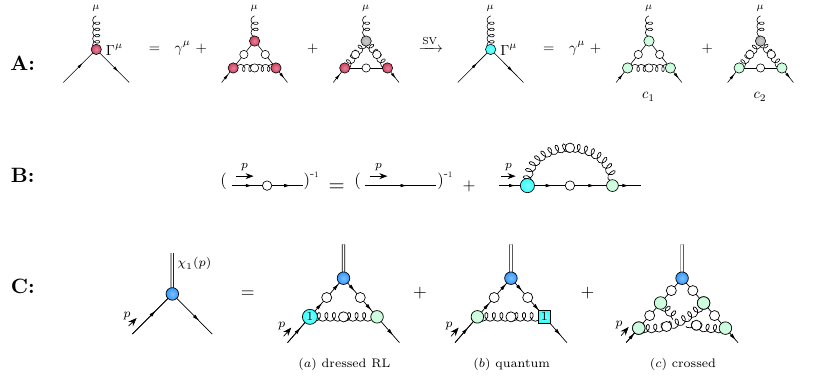}
    \caption{\textbf{\textit{Panel A:}} The SDE for the quark-gluon vertex in the SV approximation. \textbf{\textit{Panel B:}} The corresponding gap equation. \textbf{\textit{Panel C:}} The final pion BSE, after renormalization. The index ``1" 
    indicates the part of the cyan vertex that 
    contains an odd number of Dirac $\gamma$ 
    matrices, while the square vertex stands for its ``quantum" part, namely the sum of the graphs 
    $c_1$ and $c_2$.}
    \label{fig:gapeq}
\end{figure}

\begin{figure}
    \hspace*{-0.75cm}
    \includegraphics[scale=1]{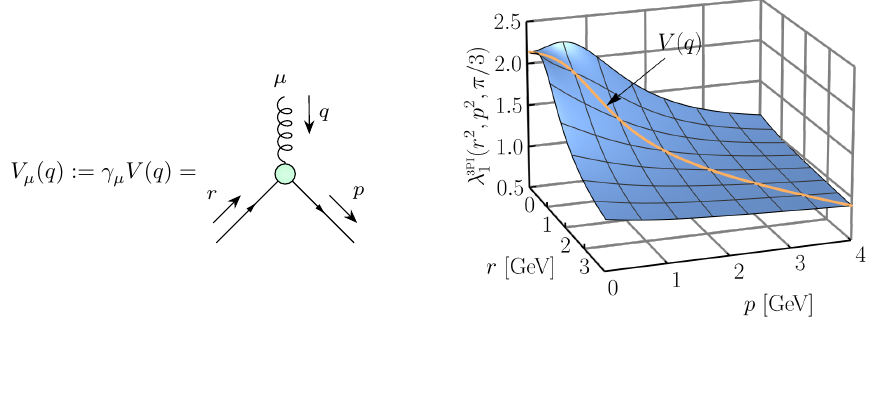}
    \caption{\textbf{\textit{Left panel}}: 
    Schematic representation of the SV approximation. \textbf{\textit{Right panel:}} 
    The classical form factor $\lambda_1$ obtained in \cite{Aguilar:2024ciu}; the orange line indicates the symmetric configuration $r^2=p^2=q^2$, used for $V(q)$. }
    \label{fig:vinput}
\end{figure}

As demonstrated in~\cite{Miramontes:2025imd},   
the dynamical equation that determines $G_5^{\mu\nu}$ 
comprises two classes of diagrams: those 
that contain $G_5^{\mu\nu}$,  and those 
where the $G_5^{\mu\nu}$ is replaced by 
$\ga^{\mu}$.  
It turns out that a symmetry-preserving truncation may be implemented at this point,
by keeping only the terms involving the 
$\ga^\mu$, provided that 
a compensating modification takes place at the 
level of the quark-gluon vertex. 
Specifically, in the diagrams retained, shown in  \figB{fig:sdeg5}, 
one must implement the replacement 
\mbox{$\g_\mu(q,r,-p) \to V_\mu(q) = \gamma_\mu V(q)$},
see \fig{fig:vinput}. 
This same replacement must be carried out  
in the defining diagrams of the SDE 
for $\g_\mu(q,r,-p)$ \cite{Alkofer:2008tt,Williams:2015cvx,Gao:2021wun}, which acquires the form 
shown in \figA{fig:gapeq}.
Note that the 
WTIs are preserved provided that 
$V(q)$ be function of 
a single kinematic variable, 
namely the momentum carried by the gluon.
In practice, the $V(q)$ chosen coincides 
with the ``symmetric limit'', $q^2=r^2=p^2$, 
of the form factor $\lambda_1$ associated with the 
classical tensor $\gamma_{\mu}$. Because 
of this particular characteristic, we coin this 
truncation as the ``{\it symmetric-vertex}`` (SV) approximation.

 We emphasize that 
the output of the SDE in \figA{fig:gapeq}, 
represented by the cyan circle, 
displays the {\it full} kinematic 
structure associated with a quark-gluon vertex,
namely eight tensorial structures (Landau gauge).
The corresponding form factors, $\lambda_i$, 
depend on three kinematic variables, \eg $q^2$, $r^2$ and $q\cdot r$. 

The new pion BSE consists of three 
diagrams, shown in 
\figC{fig:gapeq}. The first corresponds to the 
dressed version of the standard rainbow-ladder (RL)
contribution \cite{Maris:1997hd,Maris:1999bh,Eichmann:2025wgs},
while the remaining two guarantee the 
symmetry-preserving nature of the approach.

\section{Numerical analysis}
The system of coupled equations shown in \fig{fig:gapeq}
(quark-gluon vertex SDE, quark gap equation, and pion BSE) is solved using state-of-the-art 
ingredients for the correlation functions entering 
in the corresponding kernels. In particular, we 
use the gluon propagator obtained from the lattice 
simulation of \cite{Aguilar:2021okw}. As explained in \cite{Ferreira:2025anh,Ferreira:2025tzo}, the key dynamics associated 
with the emergence of a gluon mass gap 
are due to the operation of the 
Schwinger mechanism in QCD. 
For the 
the three-gluon vertex, $\Gamma_{\alpha\beta\gamma}(q,r,p)$, entering in graph $c_2$ 
of \figA{fig:gapeq}, we capitalize on the  key 
property of the 
``planar-degeneracy'' \cite{Eichmann:2014xya,Pinto-Gomez:2023lbz}: the tranversely projected $\Gamma_{\alpha\beta\gamma}(q,r,p)$ is very accurately described 
by the tree-level structure, multiplied by a special form factor
$L_{sg}(s)$. 
An accurate fit for $L_{sg}(s)$ is given in Eq.~(A1) of \cite{Aguilar:2023mam}.

\begin{figure}[!t]
    \centering
    \includegraphics[width=\linewidth]{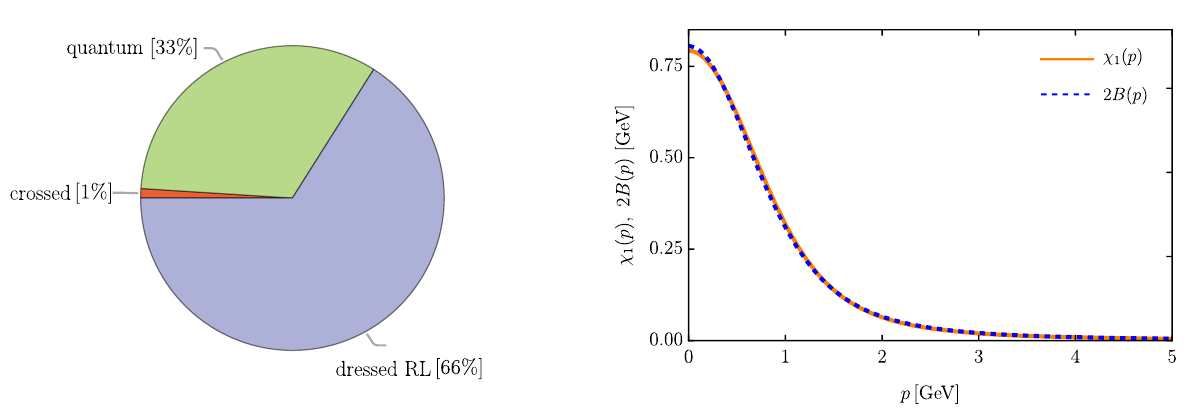}
    \caption{\textbf{\textit{Left panel}}: Contributions to the eigenvalue of the pion BSE,
    from each of the three diagrams shown in \figC{fig:gapeq}. 
\textbf{\textit{Right panel}}: Numerical confirmation of the WTI-imposed relation $\chi_1(p)=2B(p)$.}
\label{fig:pie_chart}
\end{figure}

The two main highlights of the numerical analysis are presented in \fig{fig:pie_chart}. On the left panel,
we show the individual contribution of each of the three 
BSE diagrams to the eigenvalue $\lambda(P^2)$ of the BSE,
for which, at $P^2=0$ (massless pion) we have 
$\lambda(0) =1$. On the right panel 
we demonstrate that the pion BS amplitude,  
$\chi_1(p)$, 
satisfies at a high degree of accuracy (better than $1\%$) the symmetry-induced relation 
given in \1eq{WTIcon}. 
This result constitutes a clear  
numerical confirmation of the 
symmetry-preserving nature of the 
entire approach.

\section{Conclusions}

We have presented a brief review of the 
symmetric vertex approximation, 
which allows the symmetry-preserving 
treatment of the pion dynamics  
using fully-dressed QCD correlation functions, 
and, in particular, fully-dressed quark-gluon 
vertices, with all form-factors dynamically 
evaluated.

\section*{Acknowledgments}
The work of A.S.M., J.M.M.C. and J.P. is funded by the Spanish MICINN grants PID2020-113334GB-I00 and PID2023-151418NB-I00, the Generalitat Valenciana grant CIPROM/2022/66, and CEX2023-001292-S by MCIU/AEI. J.M.P. is funded by the Deutsche Forschungsgemeinschaft (DFG, German Research Foundation) under Germany’s Excellence Strategy EXC 2181/1 - 390900948 (the Heidelberg STRUCTURES Excellence Cluster) and the Collaborative Research Centre SFB 1225 - 273811115 (ISOQUANT).

\bibliographystyle{polonica}
\bibliography{bibliography.bib}

\end{document}